\documentclass[conference]{IEEEtran}
\IEEEoverridecommandlockouts
\usepackage{cite}
\usepackage{amsmath,amssymb,amsfonts}
\usepackage{graphicx}
\usepackage{textcomp}
\usepackage{xcolor}
\usepackage{balance}
\usepackage{cite}
\usepackage{hyperref}
\usepackage{amsmath,amssymb,amsfonts}
\usepackage{amsmath}
\usepackage{amsthm}
\DeclareMathOperator*{\argmax}{argmax}

\usepackage{graphicx}
\usepackage{textcomp}
\usepackage{xcolor}
\usepackage{graphicx}
\usepackage{float}
\usepackage{subfigure}
\usepackage{amsmath}
\usepackage{amsfonts,amssymb}
\usepackage{mathrsfs}
\usepackage{mathtools}
\usepackage{bm}
\usepackage{multirow}
\usepackage{array}
\usepackage{amssymb}
\usepackage{amsmath}
\usepackage{cite}
\usepackage{url}
\usepackage{xcolor}
\usepackage{cite,graphicx,amsmath,amssymb}
\usepackage{subfigure}
\usepackage{fancyhdr}
\usepackage{mdwmath}
\usepackage{mdwtab}
\usepackage{caption}
\usepackage{amsthm}
\usepackage{setspace}
\usepackage{bm}
\usepackage{mathtools}
\usepackage{dsfont}
\usepackage{bbm}
\usepackage{algorithm}
\usepackage{algorithmic}

\newtheorem{theorem}{Theorem}

\newtheorem{lemma}{Lemma}

\newtheorem{corollary}{Corollary}

\makeatletter
\newcommand{\biggg}{\bBigg@{3}}
\newcommand{\Biggg}{\bBigg@{3.5}}
\makeatother
\def\BibTeX{{\rm B\kern-.05em{\sc i\kern-.025em b}\kern-.08em
    T\kern-.1667em\lower.7ex\hbox{E}\kern-.125emX}}
\begin{document}

\title{Statistical-CSI-Based Antenna Selection and Precoding in Uplink MIMO}

\author{\IEEEauthorblockN{Chongjun~Ouyang$^{\star}$, Ali~Bereyhi$^{\dag}$, Saba Asaad$^{\dag}$, Ralf~R.~M\"{u}ller$^{\dag}$, and Hongwen~Yang$^{\star}$}
$^{\star}$School of Information and Communication Engineering, Beijing University of Posts and Telecommunications\\
$^{\dag}$Institute for Digital Communications (IDC), Friedrich-Alexander-Universit\"{a}t Erlangen-N\"{u}rnberg\\
Email: $^{\star}$\{DragonAim, yanghong\}@bupt.edu.cn,~$^{\dag}$\{ali.bereyhi, saba.asaad, ralf.r.mueller\}@fau.de}

\maketitle

\begin{abstract}
Classical antenna selection schemes require instantaneous channel state information (CSI). This leads to high signaling overhead in the system. This work proposes a novel joint receive antenna selection and precoding scheme for multiuser multiple-input multiple-output uplink transmission that relies only on the long-term statistics of the CSI. The proposed scheme designs the switching network and the uplink precoders, such that the expected throughput of the system in the long term is maximized. Invoking results from the random matrix theory, we derive a closed-form expression for the expected throughput of the system. We then develop a tractable iterative algorithm to tackle the throughput maximization problem, capitalizing on the alternating optimization and majorization-maximization (MM) techniques. Numerical results substantiate the efficiency of the proposed approach and its superior performance as compared with the baseline.
\end{abstract}


\section{Introduction}
Antenna selection is a promising approach to alleviate the high radio frequency (RF) costs of massive multiple-input multiple-output (MIMO) systems \cite{Molisch2004,Asaad2018}. In this approach, only a subset of available antennas is set active in each coherence time resulting in lower RF costs without significant performance degradation, as compared with the implementation with full complexity. Extensive lines of work have emerged for antenna selection algorithm design in small-scale and large-scale MIMO systems; see \cite{Bereyhi2017,Asaad2018_JSAC,Kuai2020,Gershman2004,Gao2018,Ouyang2019,Ouyang2020_TCOM,Ali2018} and the references therein. Most existing works rely on the knowledge of instantaneous channel state information (CSI). This limits the practical implementation of the resulting selection algorithms in two respects: firstly, in practical use-cases with fast time variation of channel coefficients, the frequent tuning of the switching network and reallocation of the transmit powers based on the instantaneous CSI leads to heavy signal processing overhead. This burden is unaffordable for various practical MIMO systems. Secondly, the smaller number of RF chains as compared to the number of antennas increases the CSI acquisition overhead. For instance, in a time-division duplex (TDD) MIMO system, the base station (BS) has to recycle the available RF chains to acquire the full CSI via uplink training. This recycling procedure can extend the training duration and hence shorten the data transmission duration, leading to severe spectral efficiency losses \cite{Gao2018,Ouyang2020_TCOM,Kuai2020}.

The mentioned challenges reveals the necessity of designing antenna selection algorithms based on statistical CSI. Compared with instantaneous CSI, the statistical CSI, e.g., the spatial correlation and channel mean, changes over a considerably longer period of time. Hence, tuning of the switching network based on the statistical CSI reduces significantly the update rate. Moreover, with such algorithms, we only need to estimate the active channel in each coherence time instead of the entire channel. This reduces further the burden of CSI acquisition due to RF chains’ recycling and improves the spectral efficiency. Motivated by these enhancements, various lines of work have studied the design of antenna selection strategies based on the statistical CSI; see for instance \cite{Dai2006,Mehta2021,Lu2022}. These studies are however limited to specific settings, e.g., single-user scenarios. In this work, we develop a generic framework for antenna selection based on the statistical CSI.

\subsection{Contributions}
We investigate antenna selection-aided multiuser MIMO (MU-MIMO) uplink transmission, assuming only the statistical CSI is available at the BS. We formulate the design problem as a maximization of the ergodic sum-rate. The principle problem is NP-hard, as it deals with stochastic integer programming. We address this challenging problem through the following contributions: 1) Invoking the random matrix theory, we derive a closed form expression for the asymptotic ergodic sum-rate. We then use the asymptotic expression to reformulate the principle problem. 2) Using the water-filling, greedy search, and majorization-maximization (MM) techniques, we develop a low-complexity algorithm to approximate the optimal joint design. The analytical derivations are then validated via numerical simulations demonstrating the potential of exploiting statistical CSI to promote system performance. The results verify the capability of the proposed approach to obtain higher system throughput compared to the baseline.
\subsection{Notation}
Throughout this paper, scalars, vectors, and matrices are denoted by non-bold, bold lower-case, and bold upper-case letters, respectively. For the matrix $\mathbf A$, $[\mathbf A]_{i,j}$, ${\mathbf{A}}^{\mathsf T}$, and ${\mathbf{A}}^{\mathsf H}$ denote the $(i,j)$th entry, transpose, and transpose conjugate of $\mathbf A$, respectively. For the square matrix $\mathbf B$, ${\mathbf B}^{\frac{1}{2}}$, ${\mathbf B}^{-1}$, ${\mathsf{tr}}(\mathbf B)$. and $\det(\mathbf B)$ denote the principal square root, inverse, trace, and determinant of $\mathbf B$, respectively. The notation $[\mathbf a]_{i}$ denotes the $i$th entry of vector $\mathbf a$, and $\mathsf{diag}\{\mathbf a\}$ returns a diagonal matrix whose diagonal elements are entries of $\mathbf a$. The identity matrix, zero matrix, and all-one vector are represented by $\mathbf I$, $\mathbf 0$, and $\mathbf 1$, respectively. The matrix inequalities ${\mathbf A}\succeq{\mathbf 0}$ and ${\mathbf A}\succ{\mathbf 0}$ imply that $\mathbf A$ is positive semi-definite and positive definite, respectively. The set $\mathbbmss{C}$ stands for the complex plane and notation ${\mathbbmss{E}}[\cdot]$ represents mathematical expectation. The Hadamard product is shown by $\odot$. The notation $[K]$ represents the integer set $\{1, \ldots ,K\}$, and $\setminus$ denotes the set difference. Finally, ${\mathcal{CN}}\left({\mathbf 0},{\mathbf X}\right)$ is used to denotes the circularly-symmetric complex Gaussian distribution with mean zero and covariance matrix $\mathbf X$.

\section{System Model and Problem Formulation}
\begin{figure}[!t]
    \centering
    \subfigbottomskip=0pt
	\subfigcapskip=-5pt
    \setlength{\abovecaptionskip}{0pt}
    \includegraphics[width=0.3\textwidth]{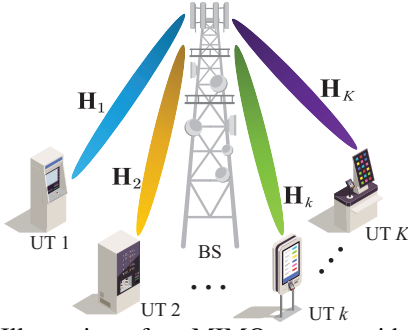}
    \caption{Illustration of an MIMO system with $K$ UTs}
    \label{system_model}
    \vspace{-20pt}
\end{figure}

An MU-MIMO setting is considered in which $K$ multiple-antenna user terminals (UTs) send messages simultaneously to an $N$-antenna BS; see {\figurename} {\ref{system_model}}. We denote the number of antennas at UT $k\in[K]$ by $N_k$, and the transmit signal conveyed by UT $k$ with ${\mathbf x}_k\in{\mathbbmss{C}}^{N_k\times1}$. The transmit signals are assumed to be zero-mean processes with covariance matrices $\{{\mathbf Q}_k={\mathbbmss E}\{{\mathbf x}_k{\mathbf x}_k^{\mathsf{H}}\}\in{\mathbbmss C}^{N_k\times N_k}\}_{k=1}^{K}$. Following the standard multiple access setting, ${\mathbf{x}}_k$ is assumed to be independent of the signals sent by other UTs, i.e., $\mathbbmss{E}\left\{{\mathbf x}_k{\mathbf x}_{k'}^{\mathsf{H}}\right\}={\mathbf 0}$ for $k\neq k'$.

In this setting, the received signal at the BS is given by
{\setlength\abovedisplayskip{2pt}
\setlength\belowdisplayskip{2pt}
\begin{align}\label{Signal_Model}
{\mathbf{y}}=\sum\nolimits_{k=1}^{K}\mathbf{H}_k{\mathbf x}_k+\mathbf{n},
\end{align}
}where $\mathbf{H}_k\in{\mathbbmss{C}}^{N\times N_k}$ is the channel matrix from UT $k$ to the BS and $\mathbf{n}$ is additive white Gaussian noise (AWGN) with mean zero and variance $\sigma^2$, i.e., ${\mathbf{n}}\sim{\mathcal{CN}}({\mathbf 0},\sigma^2{\mathbf I})$.
\subsection{Receive Antenna Selection}
The BS has $L<N$ RF chains and thus uses a switching network to select a subset of receive
antennas. This switching network connects the selected antennas to the available $L$ RF chains at the BS. As a result, the received signal at the RF front-end of the BS is given by
{\setlength\abovedisplayskip{2pt}
\setlength\belowdisplayskip{2pt}
\begin{align}
{\mathbf r}={\mathbf S}{\mathbf y}={\mathbf S}\sum\nolimits_{k=1}^{K}\mathbf{H}_k{\mathbf x}_k+{\mathbf S}{\mathbf n},
\end{align}
}where ${\mathbf S}={\{0,1\}}^{L\times N}$ is the antenna selection matrix with
{\setlength\abovedisplayskip{2pt}
\setlength\belowdisplayskip{2pt}
\begin{align}
[{\mathbf S}]_{l,n}=\begin{cases}
1& {\text{antenna}}~n~{\text{is connected to}}~{\text{RF chain}}~l\\
0& {\text{otherwise}}
\end{cases}.
\end{align}
}Since $\mathbf S$ is a rectangular permutation matrix, we have ${\mathbf{S}}{\mathbf{S}}^{{\mathsf{H}}}={\mathbf{I}}$ and ${\mathbf{S}}^{{\mathsf{H}}}{\mathbf{S}}={\mathsf{diag}}\{{\mathbf s}\}$ for ${\mathbf s}=[s_1,\ldots,s_N]^{\mathsf T}$, where $s_n\in\{0,1\}$ represents the activity of antenna $n\in[N]$, i.e., $s_n=1$ if antenna $n$ is selected, and $s_n=0$ otherwise.

\subsection{Channel Model}
To describe the UT-to-BS channel, we consider the jointly-correlated MIMO channel model (or the Weichselberger model) in the matrix of channel coefficients between UT $k$ and the BS, i.e., ${\mathbf H}_k$ is given by \cite{Lu2016}
{\setlength\abovedisplayskip{2pt}
\setlength\belowdisplayskip{2pt}
\begin{align}
{\mathbf H}_k={\mathbf U}_{{\rm R}_k}({\widetilde{\bm \Omega}}_k\odot{\widetilde{\mathbf H}}_k){\mathbf U}_{{\rm T}_k}^{\mathsf{H}},
\end{align}
}where ${\mathbf U}_{{\rm R}_k}\in{\mathbbmss C}^{N\times N}$ and ${\mathbf U}_{{\rm T}_k}\in{\mathbbmss C}^{N_k\times N_k}$ are deterministic unitary matrices, ${\widetilde{\bm \Omega}}_k\in{\mathbbmss R}^{N\times N_k}$ is a deterministic matrix with real-valued nonnegative elements, and ${\widetilde{\mathbf H}}_k\in{\mathbbmss C}^{N\times N_k}$ is independent and identically distributed (i.i.d.) standard (zero-mean and unit-variance) complex Gaussian matrix. The matrices $\{{\widetilde{\mathbf H}}_k\}_{k=1}^{K}$ are assumed to be mutually independent. In this model, ${\mathbf U}_{{\rm R}_k}$, ${\mathbf U}_{{\rm T}_k}$, and ${\widetilde{\bm \Omega}}_k$ stand for long-term statistics of the channel while ${\widetilde{\mathbf H}}_k$ captures the small scale fading. Unlike the instantaneous CSI, the statistical CSI, i.e., the spatial correlation $\{{\mathbf U}_{{\rm R}_k},{\mathbf U}_{{\rm T}_k}\}$ and the channel mean ${\bm \Omega}_k\triangleq{\widetilde{\bm \Omega}}_k\odot{\widetilde{\bm \Omega}}_k$ are rather fixed for a long period of time. It is further not a difficult task for the BS to obtain the statistical CSI through long-term feedback or covariance extrapolation.

In the sequel, we assume that the BS knows the long-term statistics for all UTs and the instantaneous CSI for the selected antennas, i.e., it has access to ${\mathcal{H}}_{\rm{s}}\triangleq\{{\mathbf U}_{{\rm R}_k},{\mathbf U}_{{\rm T}_k},{\bm\Omega}_k~{\text{for}}~k\in[K]\}$ and ${\mathcal{H}}_{\rm{i}}\triangleq\{{\mathbf S}{\mathbf H}_k~{\text{for}}~k\in[K]\}$. Specifically, at the beginning of each long period of time, the BS adjusts the switching network based on the statistical CSI; then in each coherence time, the BS estimates the instantaneous CSI of the active channel and decodes the signals sent by UTs. When the system operates in the TDD mode, the instantaneous CSI can be estimated in the uplink training phase via pilot sequences.

\subsection{Performance Metric: Ergodic Sum-Rate}
To evaluate the performance, we use the notion of ergodic sum-rate. In this respect, we discuss two types of decoding methods: joint decoding and independent decoding. Joint decoding refers to the case in which the signals of all UTs are decoded simultaneously, while independent decoding refers to the case in which the signal of each UT is decoded individually. Joint decoding is optimum yet computationally intractable, whereas independent decoding imposes lower complexity. For joint decoding, the ergodic sum-rate of the MU-MIMO system is given by
{\setlength\abovedisplayskip{2pt}
\setlength\belowdisplayskip{2pt}
\begin{align}\label{Average_Sum_Rate_Joint_Basic}
{\mathcal{R}}_{\rm{J}}={\mathbbmss E}\left\{\log\det\left({\mathbf I}+\frac{1}{\sigma^2}\sum\nolimits_{k=1}^{K}\mathbf{S}{\mathbf{H}}_k{\mathbf Q}_k{\mathbf H}_k^{\mathsf H}{\mathbf S}^{\mathsf H}\right)\right\}.
\end{align}
}The ergodic sum-rate in the case of independent decoding is given by ${\mathcal{R}}_{\rm{I}}=\sum\nolimits_{k=1}^{K}({\mathcal{R}}_{\rm{J}}-{\mathcal{R}}_{k})$, where
{\setlength\abovedisplayskip{2pt}
\setlength\belowdisplayskip{2pt}
\begin{align}\label{Average_Sum_Rate_Indep_Basic}
{\mathcal{R}}_{k}={\mathbbmss E}\left\{\!\log\det\left(\!{\mathbf I}\!+\!\frac{1}{\sigma^2}\sum\nolimits_{k'\neq k}\mathbf{S}{\mathbf{H}}_{k'}{\mathbf Q}_{k'}{\mathbf H}_{k'}^{\mathsf H}{\mathbf S}^{\mathsf H}\!\right)\!\right\}.
\end{align}
}Defining ${\bm\Delta}={\mathsf{diag}}\{s_1,\ldots,s_N\}$ and noting that ${\bm\Delta}^{\mathsf H}{\bm\Delta}={\mathbf{S}}^{{\mathsf{H}}}{\mathbf{S}}$, we can invoke Sylvester's determinant identity and rewrite \eqref{Average_Sum_Rate_Joint_Basic} and \eqref{Average_Sum_Rate_Indep_Basic} as
{\setlength\abovedisplayskip{2pt}
\setlength\belowdisplayskip{2pt}
\begin{align}
&{\mathcal{R}}_{\rm J}={\mathbbmss E}\left\{\log\det\left({\mathbf I}+\sum\nolimits_{k=1}^{K}{\bm\Delta}{\mathbf{E}}_k{{\bm\Delta}}^{\mathsf H}\right)\right\},\label{Sum_Rate_Joint_Exp2}\\
&{\mathcal{R}}_{k}={\mathbbmss E}\left\{\log\det\left({\mathbf I}+\sum\nolimits_{k'\neq k}{\bm\Delta}{\mathbf{E}}_k{{\bm\Delta}}^{\mathsf H}\right)\right\},\label{Sum_Rate_Indep_Exp2}
\end{align}
}respectively, with ${\mathbf E}_k=\frac{1}{\sigma^2}{\mathbf{H}}_k{\mathbf Q}_k{\mathbf H}_k^{\mathsf H}$ for $k\in[K]$.

\subsection{Problem Formulation}
Our ultimate goal is to find the system design that optimizes the throughput relying on the statistical channel knowledge ${\mathcal H}_{\rm{s}}$. This means that we strive to jointly design the covariance matrices ${\mathbf Q}=\{{\mathbf Q}_k\}_{k=1}^{K}$\footnote{The covariance matrices can be firstly designed at the BS side and then shared with the UTs via an error-free feedback link.} and the antenna selection vector $\mathbf{s}$, such that the ergodic sum-rate term ${\mathcal R}_{\rm{D}}$ for ${\rm D}\in\{{\rm J},{\rm I}\}$ is maximized. Hence, our design problem is characterized as
{\setlength\abovedisplayskip{2pt}
\setlength\belowdisplayskip{2pt}
\begin{equation}
\label{P_1}
\begin{split}
\max_{{\mathbf s},{\mathbf{Q}}}{\mathcal R}_{\rm D},~
{\rm{s.t.}}~&C_1:{\mathbf 1}^{\mathsf T}{\mathbf s}=L, s_n\in\{0,1\},{\text{for}}~n\in[N],\\
&C_2:{\mathsf{tr}}({\mathbf Q}_k)\leq p_k,{\mathbf Q}_k\succeq{\mathbf 0},{\text{for}}~k\in[K],
\end{split}\tag{${\mathcal P}_1$}
\end{equation}
}where $p_{k}$ is the power budget at UT $k$. It is worth mentioning that problem \eqref{P_1} is challenging due to three main reasons. Firstly, computing the expectation in \eqref{Sum_Rate_Joint_Exp2} and \eqref{Sum_Rate_Indep_Exp2} incurs a prohibitive computational cost. Secondly, the discrete constraints in $C_1$ make \eqref{P_1} essentially an NP-hard problem. Finally, the tight coupling between $\mathbf Q$ and $\mathbf s$ further complicates the optimization procedure. In the sequel, we strive to confront the above challenges and develop an efficient framework to approximate the optimal design via a feasible computational complexity.

\section{Optimization for Joint Decoding}
In \eqref{P_1}, the variables ${\mathbf Q}$ and $\mathbf s$ are nonlinearly coupled in $\mathcal{P}_1$ and are hence complicated to be optimized simultaneously. To facilitate the joint design, we resort to the alternating optimization (AO) method, i.e., we alternate between two marginal problems: optimize ${\mathbf Q}$ while treating $\mathbf s$ as fixed and optimize $\mathbf s$ while treating ${\mathbf Q}$ as fixed. In the sequel, we discuss each marginal problem separately.
\subsection{Optimizing the Transmit Covariance Matrices}
The first marginal problem is given by
{\setlength\abovedisplayskip{2pt}
\setlength\belowdisplayskip{2pt}
\begin{equation}
\label{P_2}
\begin{split}
\max_{{\mathbf{Q}}}~{\mathcal R}_{\rm J}(\mathbf Q),~{\rm{s.t.}}~C_2,
\end{split}\tag{${\mathcal P}_2$}
\end{equation}
}where
{\setlength\abovedisplayskip{2pt}
\setlength\belowdisplayskip{2pt}
\begin{align}
{\mathcal R}_{\rm J}(\mathbf Q)\!=\!{\mathbbmss E}\left\{\!\log\det\left(\!{\mathbf I}\!+\!\sum\nolimits_{k=1}^{K}\!\!{\mathbf A}_k{\mathbf U}_{{\rm T}_k}^{\mathsf{H}}{\mathbf Q}_k{\mathbf U}_{{\rm T}_k}{{\mathbf A}_k^{\mathsf H}}\!\right)\!\right\}\nonumber
\end{align}
}with ${\mathbf A}_k={\bm\Delta}{\mathbf U}_{{\rm R}_k}({\widetilde{\bm \Omega}}_k\odot{\widetilde{\mathbf H}}_k)$.

To facilitate the optimization of ${\mathbf Q}$, we decompose the input covariance as ${\mathbf Q}_k={\mathbf V}_k{\bm\Lambda}_k{\mathbf V}_k^{\mathsf H}$ identifying the eigenvectors of ${\mathbf Q}_k$ with the columns of the unitary matrix ${\mathbf V}_k$ and its eigenvalues with the diagonal entries of ${\bm\Lambda}_k={\mathsf{diag}}\{\lambda_{k,1},\ldots,\lambda_{k,N_k}\}$. It is worth mentioning that the eigenvectors indicate the directions (in vector space) on which signalling takes place while the eigenvalues signify the transmit powers allocated onto each such eigenvector.

We next invoke results from random matrix theory to derive a closed-form expression for the objective function in the asymptotic regime. We start the derivations by the following lemma:
\vspace{-5pt}
\begin{lemma}
The columns of ${\mathbf U}_{{\rm T}_k}$ give the eigenvectors of the sum-rate-optimal transmit covariance, i.e., the optimal ${\mathbf V}_k$ is identical to the unitary matrix ${\mathbf U}_{{\rm T}_k}$.
\end{lemma}
\vspace{-5pt}
\begin{IEEEproof}
The proof is similar to that provided in the proof of Theorem 1 in \cite{Tulino2006}, and thus omitted here for brevity.
\end{IEEEproof}
By setting ${\mathbf V}_k={\mathbf U}_{{\rm T}_k}$ for $k\in[K]$, the transmit covariance matrix design problem in \eqref{P_2} can be transformed to a power allocation problem as follows:
{\setlength\abovedisplayskip{2pt}
\setlength\belowdisplayskip{2pt}
\begin{equation}
\label{P_3}
\begin{split}
\max_{{\bm\Lambda}}~&{\mathcal R}_{\rm J}({\bm\Lambda})\!=\!{\mathbbmss E}\left\{\log\det\left({\mathbf I}\!+\!\sum\nolimits_{k=1}^{K}\!\!{\mathbf A}_k{\bm\Lambda}_k{{\mathbf A}_k^{\mathsf H}}\right)\right\}\\
{\rm{s.t.}}~&C_3:{\mathsf{tr}}({\bm\Lambda}_k)\leq p_k,{\bm\Lambda}_k\succeq{\mathbf 0},{\text{for}}~k\in[K],
\end{split}\tag{$\mathcal{P}_3$}
\end{equation}
}where ${\bm\Lambda}=\{{\bm\Lambda}_k\}_{k=1}^{K}$. We note that problem \eqref{P_3} is a standard convex problem that can be solved via conventional stochastic programming techniques. Yet, since ${\mathcal R}_{\rm J}({\bm\Lambda})$ involves an expectation, we need to approximate it by the Monte-Carlo method via averaging over a large number of samples, which is computationally expensive. To avoid channel averaging, we approximate the expectation with its large-system limit using tools from the random matrix theory:
\vspace{-5pt}
\begin{lemma}\label{Lemma_DE}
Let $N$ and $N_k$ for $k\in[K]$ both tend to infinity with the ratios $c_k=N_k/N$ fixed. Given ${\bm\Lambda}$ and $\mathbf s$, the ergodic sum-rate with joint decoding is asymptotically approximated by
{\setlength\abovedisplayskip{2pt}
\setlength\belowdisplayskip{2pt}
\begin{equation}
\begin{split}
\mathcal{R}_{\rm{J}}\approx{\dot{\mathcal{R}}}_{\rm J}&=\sum\nolimits_{k=1}^{K}\log\det({\mathbf{I}}+{\bm\Xi}_k{\bm\Lambda}_k)\\
&+\log\det({\mathbf{I}}+{\mathbf R})
-\sum\nolimits_{k=1}^{K}{\bm\gamma}_k^{\mathsf T}{\bm\Omega}_k{\bm\psi}_k,\label{DE_Sum_Rate}
\end{split}
\end{equation}
}where ${\bm\gamma}_k=[\gamma_{k,1},\ldots,\gamma_{k,N}]^{\mathsf T}$, ${\bm\psi}_k=[\psi_{k,1},\ldots,\psi_{k,N_k}]^{\mathsf T}$,
{\setlength\abovedisplayskip{2pt}
\setlength\belowdisplayskip{2pt}
\begin{align}
&{\bm\Xi}_k={\mathsf{diag}}\{{\bm\Omega}_k^{\mathsf T}{\bm\gamma}_k\},\\
&{\mathbf R}=\frac{1}{\sigma^{2}}\sum\nolimits_{k=1}^{K}{\bm\Delta}{\mathbf U}_{{\rm R}_k}{\mathsf{diag}}\{{\bm\Omega}_k{\bm\psi}_k\}{\mathbf U}_{{\rm R}_k}^{\mathsf H}{\bm\Delta}^{\mathsf H}.
\end{align}
}The auxiliary quantities ${\bm\gamma}=\{\gamma_{k,n}\}_{k\in[K],n\in[N]}$ and ${\bm\psi}=\{\psi_{k,m}\}_{k\in[K],m\in[N_k]}$ are the unique solutions
to the following iterative equations:
{\setlength\abovedisplayskip{2pt}
\setlength\belowdisplayskip{2pt}
\begin{align}
&\gamma_{k,n}=\frac{1}{\sigma^{2}}{\mathbf u}_{k,n}^{\mathsf H}({\mathbf I}+{\mathbf R})^{-1}{\mathbf u}_{k,n},\label{iterative_equation_1}\\
&\psi_{k,m}={\lambda_{k,m}}/({1+\lambda_{k,m}\xi_{k,m}}),\label{iterative_equation_2}
\end{align}
}where ${\mathbf u}_{k,n}$ denotes the $n$th column of ${\bm\Delta}{\mathbf U}_{{\rm R}_k}$, and $\xi_{k,m}$ is the $m$th diagonal element of ${\bm\Xi}_k$. Given an initial point of ${\bm\gamma}$ or ${\bm\psi}$, the fixed-point solutions of ${\bm\gamma}$ and ${\bm\psi}$ can be solved via cyclically updating them by \eqref{iterative_equation_1} and \eqref{iterative_equation_2}.
\end{lemma}
\vspace{-5pt}
\begin{IEEEproof}
The proof follows the results in \cite{Couillet2011}.
\end{IEEEproof}
Note that the adopted asymptotic approximation is sufficiently accurate for small-scale MIMO systems. We also note that ${\dot{\mathcal{R}}}_{\rm J}$ is solely determined by the statistical channel knowledge ${\mathcal H}_{\rm{s}}$. By replacing ${\mathcal{R}}_{\rm J}$ in \eqref{P_1} with its asymptotic approximation ${\dot{\mathcal{R}}}_{\rm J}$, we can obtain the asymptotic optimization problem as follows:
{\setlength\abovedisplayskip{2pt}
\setlength\belowdisplayskip{2pt}
\begin{equation}\label{P_4}
\max\nolimits_{{\mathbf s},{\bm\Lambda}}~\dot{\mathcal{R}}_{\rm J},~{\rm{s.t.}}~C_1,C_3.\tag{${\mathcal P}_4$}
\end{equation}
}The above problem enables us to design $\bm\Lambda$ and $\mathbf{s}$ by exploiting the statistical knowledge of the channels without knowing the actual channel realization. It is emphasized that $\bm\psi$ and $\bm\gamma$ are both functions of $({\bm\Lambda},\mathbf{s})$, yet lacking any explicit expressions. We hence resort to the AO method by updating $({\bm\Lambda},\mathbf{s})$ and $\{{\bm\psi},{\bm\gamma}\}$ separately in an iterative fashion, as in \cite{Couillet2011}.

Given $\{{\bm\psi},{\bm\gamma},\mathbf{s}\}$, the marginal optimization for $\{{\bm\Lambda}_k\}_{k=1}^{K}$ reduces to $K$ subproblems with the $k$th one being
{\setlength\abovedisplayskip{2pt}
\setlength\belowdisplayskip{2pt}
\begin{align}\label{P_5}
{\bm\Lambda}_k^{\star}=\argmax\nolimits_{{\mathsf{tr}}({\bm\Lambda}_k)\leq p_k}{\log\det({\mathbf{I}}+{\bm\Xi}_k{\bm\Lambda}_k)},\tag{${\mathcal P}_{5,k}$}
\end{align}
}where ${\bm\Xi}_k={\mathsf{diag}}\{\xi_{k,1},\ldots,\xi_{k,N_k}\}\succeq{\mathbf 0}$ \cite{Couillet2011} and ${\bm\Lambda}_k^{\star}\triangleq{\mathsf{diag}}\{\lambda_{k,1}^{\star},\ldots,\lambda_{k,N_k}^{\star}\}$. Consequently, the optimal power allocation matrix of UT $k$ is given by the water-filling with respective to the channel covariance matrix ${\bm\Xi}_k$, i.e.,
{\setlength\abovedisplayskip{2pt}
\setlength\belowdisplayskip{2pt}
\begin{align}
\lambda_{k,m}^{\star}=\max\{{1}/{\mu_k}-{1}/{\xi_{k,m}},0\}\label{Power_Allocation_Design}
\end{align}
}with $\mu_k$ chosen to satisfy the power constraint ${\mathsf{tr}}({\bm\Lambda}_k^{\star})= p_k$.
\subsection{Optimizing the Antenna Selection Vector}
We now consider the second marginal problem in which the objective is maximized over the antenna selection vector $\mathbf s$ while treating UTs' power allocation matrices $\bm\Lambda$ and auxiliary variables $\{{\bm\gamma},{\bm\psi}\}$ as fixed variables. Dropping all the constant terms in \eqref{DE_Sum_Rate}, we mathematically express the optimization over $\mathbf s$ as follows:
{\setlength\abovedisplayskip{2pt}
\setlength\belowdisplayskip{2pt}
\begin{equation}\label{P_6}
\max_{{\mathbf s}}~\log\det\left({\mathbf{I}}+{\bm\Delta}{\mathbf B}{\bm\Delta}^{\mathsf H}\right),~{\rm{s.t.}}~C_1,\tag{${\mathcal P}_6$}
\end{equation}
}where \cite{Couillet2011}
{\setlength\abovedisplayskip{2pt}
\setlength\belowdisplayskip{2pt}
\begin{equation}
{\mathbf B}=\frac{1}{\sigma^2}\sum\nolimits_{k=1}^{K}{\mathbf U}_{{\rm R}_k}{\mathsf{diag}}\{{\bm\Omega}_k{\bm\psi}_k\}{\mathbf U}_{{\rm R}_k}^{\mathsf H}\succeq{\mathbf 0}.
\end{equation}
}We now regard the objective of \eqref{P_6} as the data rate of a hypothetical MIMO system with receive antenna selection, which is described by ${\mathbf y}_{\rm{h}}={\bm\Delta}{\mathbf B}^{1/2}{\mathbf s}_{\rm h}+{\mathbf n}_{\rm h}$. Specifically, $\bm\Delta$, ${\mathbf B}^{1/2}$, ${\mathbf s}_{\rm h}\sim{\mathcal{CN}}\left({\mathbf 0},{\mathbf I}\right)$, and ${\mathbf n}_{\rm h}\sim{\mathcal{CN}}\left({\mathbf 0},{\mathbf I}\right)$ represent the hypothetical antenna selection matrix, channel matrix, signal symbol vector, and AWGN, respectively. Mathematically, the selection matrix $\bm\Delta$ picks $L$ proper rows from the matrix ${\mathbf B}^{1/2}$.

The solution of problem \eqref{P_6} can be found via an exhaustive search or a branch-and-bound search \cite{Gao2018}, both being computationally cumbersome. For further simplifications, we resort to the low-complexity greedy search with $L$ steps, where at the $l$th step, the $J_l$th row of ${\mathbf B}^{1/2}$ is selected from the candidate set $\mathcal{B}_l$ to maximize the rate increment \cite{Gershman2004}. More precisely, let ${\mathbf b}_{n}^{\mathsf{T}}$ denote the $n$th row of ${\mathbf B}^{1/2}$ and ${\mathbf B}_l$ denote the sub-matrix of ${\mathbf B}^{1/2}$ formed by the $l$ selected rows. Following the principle of the greedy search, we can obtain
{\setlength\abovedisplayskip{2pt}
\setlength\belowdisplayskip{2pt}
\begin{align}
J_l=\argmax\nolimits_{l'\in{\mathcal B}_l}({\mathcal{R}}_{l-1,l'}-{\mathcal{R}}_{l-1})~{\text{for}}~l\in[L],\label{Greedy_Search_Procedure}
\end{align}
}where ${\mathcal{R}}_{l-1}=\log\det({\mathbf I}+{\mathbf B}_{l-1}^{\mathsf{H}}{\mathbf B}_{l-1})$, ${\mathcal{R}}_{l-1,l'}=\log\det({\mathbf I}+{\mathbf B}_{l-1}^{\mathsf{H}}{\mathbf B}_{l-1}+{\mathbf b}_{l'}{\mathbf b}_{l'}^{\mathsf H})$, and ${\mathcal B}_l=[N]\setminus\{J_1,\ldots,J_{l-1}\}$ with $\mathcal{R}_0=0$ and ${\mathcal B}_1=\varnothing$. Using Sylvester’s determinant identity, we calculate the rate increment as
{\setlength\abovedisplayskip{2pt}
\setlength\belowdisplayskip{2pt}
\begin{align}
{\Delta}_{l-1,l'}={\mathcal{R}}_{l-1,l'}-{\mathcal{R}}_{l-1}=\log(1+{\mathbf b}_{l'}^{\mathsf H}{\mathbf G}_{l-1}{\mathbf b}_{l'}),
\end{align}
}where ${\mathbf G}_{l-1}=({\mathbf I}+{\mathbf B}_{l-1}^{\mathsf{H}}{\mathbf B}_{l-1})^{-1}$ with ${\mathbf G}_{0}={\mathbf I}$. We further invoke the Woodbury formula and show that ${\mathbf G}_l$ satisfies the following recursive equality:
{\setlength\abovedisplayskip{2pt}
\setlength\belowdisplayskip{2pt}
\begin{align}
{\mathbf G}_l=({\mathbf{G}}_{l-1}^{-1}+{\mathbf b}_{J_{l-1}}{\mathbf b}_{J_{l-1}}^{\mathsf H})^{-1}={\mathbf{G}}_{l-1}-{\mathbf{g}}_{l-1}{\mathbf{g}}_{l-1}^{\mathsf H},\label{Help_Greedy_Matrix}
\end{align}
}where ${\mathbf{g}}_{l-1}=\frac{{\mathbf G}_{l-1}{\mathbf b}_{J_{l-1}}}{\sqrt{1+{\mathbf b}_{J_{l-1}}^{\mathsf H}{{\mathbf G}_{l-1}}{\mathbf b}_{J_{l-1}}}}$, ${\mathbf b}_{J_{0}}={\mathbf 0}$, and ${\mathbf g}_{0}={\mathbf 0}$. Since ${\mathbf G}_0={\mathbf I}$, we can use \eqref{Help_Greedy_Matrix} to update ${\mathbf G}_l$ without computing the matrix inversion in $({\mathbf I}+{\mathbf B}_{l}^{\mathsf{H}}{\mathbf B}_{l})^{-1}$.

Taking the above steps together, we conclude a greedy algorithm that is represented in Algorithm \ref{Algorithm1}.
\begin{algorithm}[!t]
  \caption{Greedy search for solving problem $\mathcal P_6$}
  \label{Algorithm1}
  \begin{algorithmic}[1]
    \STATE Initialize $l=0$, ${\mathbf G}_0={\mathbf I}$, ${\mathbf g}_{0}={\mathbf 0}$, ${\mathcal B}_0=[N]$, and $J_{0}\!=\!\varnothing$
    \REPEAT
    \STATE Set $l=l+1$ and update ${\mathbf G}_l$ based on \eqref{Help_Greedy_Matrix}
    \STATE Update ${\mathcal B}_l={\mathcal B}_{l-1}\setminus\{J_{l-1}\}$
    \STATE Update $J_l$ based on \eqref{Greedy_Search_Procedure}
    \UNTIL{$l=L$}
  \end{algorithmic}
\end{algorithm}

\section{Optimization for Independent Decoding}
We next consider the joint design problem for the case of independent decoding. Similar to the case with joint decoding, we start the analysis by deriving an asymptotic approximation for the ergodic sum-rate in this case. We then use the derived approximation to address the design problem via AO.
\subsection{Large-System Sum-Rate}
Lemma \ref{Lemma_DE2} characterizes the ergodic sum-rate ${\mathcal R}_{\rm{I}}$ with independent decoding.
\vspace{-5pt}
\begin{lemma}\label{Lemma_DE2}
Let $N$ and $N_k$ for $k\in[K]$ both tend to infinity with the ratios $c_k=N_k/N$ fixed. Given ${\bm\Lambda}$ and $\mathbf s$, the ergodic sum-rate of the uplink MIMO channel \eqref{Signal_Model} under independent decoding can be asymptotically approximated by
{\setlength\abovedisplayskip{2pt}
\setlength\belowdisplayskip{2pt}
\begin{align}
&\mathcal{R}_{\rm{I}}\approx{\dot{\mathcal{R}}}_{\rm I}=K{\dot{\mathcal{R}}}_{0}-\sum\nolimits_{k=1}^{K}\log\det({\mathbf{I}}+\overline{\mathbf R}_k)
-\sum\nolimits_{k=1}^{K}\nonumber\\
&\times\sum\nolimits_{k'\neq k}(\log\det({\mathbf{I}}+\overline{\bm\Xi}_{k,k'}{\mathbf Q}_{k'})
-\overline{\bm\gamma}_{k,k'}^{\mathsf T}{\bm\Omega}_{k'}\overline{\bm\psi}_{k,k'}),
\end{align}
}where $\overline{\bm\gamma}_{k,k'}=[\overline{\gamma}_{k,k',1},\ldots,\overline{\gamma}_{k,k',N}]^{\mathsf T}$, $\overline{\bm\psi}_{k,k'}=[\overline{\psi}_{k,k',1},\ldots,\overline{\psi}_{k,k',N_{k'}}]^{\mathsf T}$,
{\setlength\abovedisplayskip{2pt}
\setlength\belowdisplayskip{2pt}
\begin{align}
&\overline{\bm\Xi}_{k,k'}={\mathbf U}_{{\rm T}_{k'}}{\mathsf{diag}}\{{\bm\Omega}_{k'}^{\mathsf T}\overline{\bm\gamma}_{k,k'}\}{\mathbf U}_{{\rm T}_{k'}^{\mathsf H}},\\
&\overline{\mathbf R}_k=\frac{1}{\sigma^{2}}\sum\nolimits_{k'\neq k}{\bm\Delta}{\mathbf U}_{{\rm R}_{k'}}{\mathsf{diag}}\{{\bm\Omega}_{k'}\overline{\bm\psi}_{k,k'}\}{\mathbf U}_{{\rm R}_{k'}}^{\mathsf H}{\bm\Delta}^{\mathsf H}.
\end{align}
}The auxiliary quantities $\overline{\bm\gamma}=\{\overline{\gamma}_{k,k',n}\}_{k\in[K],k'\neq k,n\in[N]}$ and $\overline{\bm\psi}=\{\overline{\psi}_{k,k',m}\}_{k\in[K],k'\neq k,m\in[N_{k'}]}$ are the unique solutions
to the following iterative equations:
{\setlength\abovedisplayskip{2pt}
\setlength\belowdisplayskip{2pt}
\begin{align}
&\overline{\gamma}_{k,k',n}={\sigma^{-2}}{\mathbf u}_{k',n}^{\mathsf H}({\mathbf I}+\overline{\mathbf R}_k)^{-1}{\mathbf u}_{k',n},\\
&\overline{\psi}_{k,k',m}={\mathbf v}_{k',m}^{\mathsf H}{\mathbf Q}_{k'}({\mathbf I}+\overline{\bm\Xi}_{k,k'}{\mathbf Q}_{k'})^{-1}{\mathbf v}_{k',m},
\end{align}
}where ${\mathbf v}_{k',m}$ denotes the $m$th column of ${\mathbf U}_{{\rm T}_{k'}}$. Moreover, ${\dot{\mathcal{R}}}_{0}$ is given by
{\setlength\abovedisplayskip{2pt}
\setlength\belowdisplayskip{2pt}
\begin{equation}
\begin{split}
{\dot{\mathcal{R}}}_{0}&=\sum\nolimits_{k=1}^{K}\log\det({\mathbf{I}}+\hat{\bm\Xi}_k{\mathbf Q}_k)\\
&+\log\det({\mathbf{I}}+\hat{\mathbf R})
-\sum\nolimits_{k=1}^{K}\hat{\bm\gamma}_k^{\mathsf T}{\bm\Omega}_k\hat{\bm\psi}_k,\label{DE_Sum_Rate}
\end{split}
\end{equation}
}where $\hat{\bm\gamma}_k=[\hat{\gamma}_{k,1},\ldots,\hat{\gamma}_{k,N}]^{\mathsf T}$, $\hat{\bm\psi}_k=[\hat{\psi}_{k,1},\ldots,\hat{\psi}_{k,N_k}]^{\mathsf T}$,
{\setlength\abovedisplayskip{2pt}
\setlength\belowdisplayskip{2pt}
\begin{align}
&\hat{\bm\Xi}_k={\mathbf U}_{{\rm T}_{k}}{\mathsf{diag}}\{{\bm\Omega}_{k}^{\mathsf T}\hat{\bm\gamma}_{k}\}{\mathbf U}_{{\rm T}_{k}}^{\mathsf H},\\
&\hat{\mathbf R}={\sigma^{-2}}\sum\nolimits_{k=1}^{K}{\bm\Delta}{\mathbf U}_{{\rm R}_k}{\mathsf{diag}}\{{\bm\Omega}_k\hat{\bm\psi}_k\}{\mathbf U}_{{\rm R}_k}^{\mathsf H}{\bm\Delta}^{\mathsf H}.
\end{align}
}The auxiliary quantities $\hat{\bm\gamma}=\{\hat{\gamma}_{k,n}\}_{k\in[K],n\in[N]}$ and $\hat{\bm\psi}=\{\hat{\psi}_{k,m}\}_{k\in[K],m\in[N_k]}$ are the unique solutions
to the following iterative equations:
{\setlength\abovedisplayskip{2pt}
\setlength\belowdisplayskip{2pt}
\begin{align}
&\hat{\gamma}_{k,n}={\sigma^{-2}}{\mathbf u}_{k,n}^{\mathsf H}({\mathbf I}+\hat{\mathbf R})^{-1}{\mathbf u}_{k,n},\\
&\hat{\psi}_{k,m}={\mathbf v}_{k,m}^{\mathsf H}{\mathbf Q}_{k}({\mathbf I}+\hat{\bm\Xi}_{k}{\mathbf Q}_{k})^{-1}{\mathbf v}_{k,m}.
\end{align}
}\end{lemma}
\vspace{-5pt}
\begin{IEEEproof}
Similar to the proof of Lemma \ref{Lemma_DE}.
\end{IEEEproof}
By replacing the objective function ${\mathcal R}_{\rm I}$ in \eqref{P_1} with ${\dot{\mathcal{R}}}_{\rm I}$, we reformulate problem \eqref{P_1} as its asymptotic version as
{\setlength\abovedisplayskip{2pt}
\setlength\belowdisplayskip{2pt}
\begin{align}\label{P_7}
\max\nolimits_{{\mathbf s},{\mathbf Q}}~&\dot{\mathcal{R}}_{\rm I},~{\rm{s.t.}}~C_1,C_2.\tag{${\mathcal P}_7$}
\end{align}
}We then exploit the AO method to solve problem \eqref{P_7} by updating ${\mathbf Q}$, $\mathbf{s}$, and $\{\hat{\bm\psi},\overline{\bm\psi},\hat{\bm\gamma},\overline{\bm\gamma}\}$ in an alternating manner.
\subsection{Optimizing the Transmit Covariance Matrices}
Given $\{\hat{\bm\psi},\overline{\bm\psi},\hat{\bm\gamma},\overline{\bm\gamma},{\mathbf s}\}$, the marginal optimization for $\{{\mathbf Q}_k\}_{k=1}^{K}$ reduces to $K$ subproblems with the $k$th one being
{\setlength\abovedisplayskip{2pt}
\setlength\belowdisplayskip{2pt}
\begin{align}\label{Independent_Decoding_Transmit_Covariance}
{\mathbf Q}_k^{\star}=\argmax\nolimits_{{\mathbf Q}_k}(f_{1,k}^{+}({\mathbf Q}_k)-f_{1,k}^{-}({\mathbf Q}_k)),\tag{${\mathcal P}_{8,k}$}
\end{align}
}where ${\mathsf{tr}}({\mathbf Q}_k)\leq p_k$ with ${\mathbf Q}_k\succeq{\mathbf 0}$, and
{\setlength\abovedisplayskip{2pt}
\setlength\belowdisplayskip{2pt}
\begin{align}
f_{1,k}^{+}({\mathbf Q}_k)&=K\log\det({\mathbf{I}}+\hat{\bm\Xi}_k{\mathbf Q}_k),\\
f_{1,k}^{-}({\mathbf Q}_k)&=\sum\nolimits_{k'\neq k}\log\det({\mathbf{I}}+\overline{\bm\Xi}_{k',k}{\mathbf Q}_{k}),
\end{align}
}with $\hat{\bm\Xi}_k\succeq{\mathbf 0}$ and $\overline{\bm\Xi}_{k',k}\succeq{\mathbf 0}$ for $k'\neq k$ \cite{Couillet2011}. The objective of problem \eqref{Independent_Decoding_Transmit_Covariance} is in the form of difference of two concave functions, $f_{1,k}^{+}({\mathbf Q}_k)$ and $f_{1,k}^{-}({\mathbf Q}_k)$. We therefore exploit the iterative MM-based method \cite{Sun2017} to obtain a suboptimal solution of problem \eqref{Independent_Decoding_Transmit_Covariance}. Let ${\mathbf Q}_k^{(j)}$ be an initial feasible solution of \eqref{Independent_Decoding_Transmit_Covariance} at the $j$th iteration. Using first-order Taylor expansion, we obtain a convex upper bound for $f_{1,k}^{-}({\mathbf Q}_k)$:
{\setlength\abovedisplayskip{2pt}
\setlength\belowdisplayskip{2pt}
\begin{align}
f_{1,k}^{-}({\mathbf Q}_k)\leq \Re\{{\mathsf{tr}}({\bm\Delta}_j^{\mathsf H}({\mathbf Q}_k-{\mathbf Q}_k^{(j)}))\}+f_{1,k}^{-}({\mathbf Q}_k^{(j)}),
\end{align}
}where ${\bm\Delta}_j=\sum_{k'\neq k}\overline{\bm\Xi}_{k',k}^{\frac{1}{2}}({\mathbf{I}}+\overline{\bm\Xi}_{k',k}^{\frac{1}{2}}{\mathbf Q}_k^{(j)}\overline{\bm\Xi}_{k',k}^{\frac{1}{2}})^{-1}\overline{\bm\Xi}_{k',k}^{\frac{1}{2}}$. Then, for a given ${\mathbf Q}_k^{(j)}$, by replacing the non-convex $f_{1,k}^{-}({\mathbf Q}_k)$ with its convex upper bound, we transform problem \eqref{Independent_Decoding_Transmit_Covariance} into the optimization problem as follows
{\setlength\abovedisplayskip{2pt}
\setlength\belowdisplayskip{2pt}
\begin{equation}\label{P_8}
\begin{split}
\max\nolimits_{{\mathbf Q}_k}~&f_{1,k}^{+}({\mathbf Q}_k)-
\Re\{{\mathsf{tr}}({\bm\Delta}_j^{\mathsf H}{\mathbf Q}_k)\}\\
{\rm{s.t.}}~&{\mathsf{tr}}({\mathbf Q}_k)\leq p_k, {\mathbf Q}_k\succeq{\mathbf 0}.
\end{split}\tag{${\mathcal P}_{9,k}^j$}
\end{equation}
}The relaxed problem \eqref{P_8} is convex. We hence handle it via standard convex problem solvers such as CVX \cite{Boyd2014}. The proposed MM-based algorithm for solving problem \eqref{Independent_Decoding_Transmit_Covariance} is summarized in Algorithm \ref{Algorithm2}. The objective function is monotonically non-decreasing after each iteration and a stationary point of problem \eqref{Independent_Decoding_Transmit_Covariance} can be obtained \cite{Sun2017}.
\begin{algorithm}[!t]
  \caption{Proposed algorithm for solving problem \eqref{Independent_Decoding_Transmit_Covariance}}
  \label{Algorithm2}
  \begin{algorithmic}[1]
    \STATE Initialize feasible ${\mathbf Q}_k^{(0)}$, $\forall k\in{\mathcal K}$ and iteration index $j\!=\!0$
    \REPEAT
    \STATE Update ${\mathbf Q}_k^{(j+1)}$ by solving problem \eqref{P_8};
    \STATE Set $j=j+1$
    \UNTIL{convergence}
  \end{algorithmic}
\end{algorithm}
\subsection{Optimizing the Antenna Selection Vector}
The marginal problem for $\mathbf s$ is given by
{\setlength\abovedisplayskip{2pt}
\setlength\belowdisplayskip{2pt}
\begin{align}\label{P_10}
\max\nolimits_{{\mathbf s}}~f_2(\mathbf{s})=f_2^{+}(\mathbf{s})-f_2^{-}(\mathbf{s}),~{\rm{s.t.}}~C_1\tag{${\mathcal P}_{10}$}.
\end{align}
}The terms appearing in \eqref{P_10} are defined as follows:
\begin{itemize}
  \item $f_2^{+}(\mathbf{s})=K\log\det({\mathbf{I}}+{\bm\Delta}\hat{\mathbf B}{\bm\Delta}^{\mathsf H})$.
  \item $\hat{\mathbf B}=\frac{1}{\sigma^{2}}\sum\nolimits_{k=1}^{K}{\mathbf U}_{{\rm R}_k}{\mathsf{diag}}\{{\bm\Omega}_k\hat{\bm\psi}_k\}{\mathbf U}_{{\rm R}_k}^{\mathsf H}\succeq{\mathbf 0}$.
  \item $f_2^{-}(\mathbf{s})=\sum\nolimits_{k=1}^{K}\log\det({\mathbf{I}}+{\bm\Delta}\overline{\mathbf B}_k{\bm\Delta}^{\mathsf H})$.
  \item $\overline{\mathbf B}_k=\frac{1}{\sigma^{2}}\sum\nolimits_{k'\neq k}{\mathbf U}_{{\rm R}_{k'}}{\mathsf{diag}}\{{\bm\Omega}_{k'}\overline{\bm\psi}_{k,k'}\}{\mathbf U}_{{\rm R}_{k'}}^{\mathsf H}\succeq{\mathbf 0}$.
\end{itemize}
The solution of problem \eqref{P_10} can be found by method of exhaustive search or branch-and-bound search \cite{Ouyang2019}, yet both being computationally cumbersome. For simplifications, we resort to the greedy search-based method to find a suboptimal solution of $\mathbf s$. The detailed steps are similar to those outlined in Algorithm \ref{Algorithm1} and are skipped here for brevity.

\section{Convergence and Complexity}
Combining the proposed methods for finding the solutions of $\bm\Lambda$ (or $\mathbf Q$) and $\mathbf s$, we reach a complete throughput maximization approach relying on statistical CSI and summarize this approach in Algorithm \ref{Algorithm3}. Similar to \cite{Couillet2011}, it is difficult to prove the convergence of the AO method adopted in Algorithm \ref{Algorithm3}. However, extensive simulations suggest that convergence is always attained.

Let $I_{\rm{AO}}$ denote the number of required iterations of Algorithm \ref{Algorithm3}. The per-iteration complexity of Algorithm \ref{Algorithm3} mainly originates from updating variables $\{{\bm\Lambda},{\mathbf s}\}$ or $\{{\mathbf Q},{\mathbf s}\}$. For joint decoding, the complexity of marginal optimizations in terms of $\bm\Lambda$ and $\mathbf s$ scales with ${\mathcal O}\left(\sum_{k=1}^{K}N_k\right)$ and ${\mathcal{O}}(LN^2)$, respectively. Hence, the overall complexity of Algorithm \ref{Algorithm3} under joint decoding scales with ${\mathcal O}(I_{\rm{AO}}(LN^2+\sum_{k=1}^{K}N_k))$, which is of a polynomial order. In the case of independent decoding, we update $\mathbf Q$ using Algorithm \ref{Algorithm2} that requires a total of $I_{\rm{MM}}$ iterations. As the relaxed problem \eqref{P_8} belongs to the semidefinite program, the complexity of solving this problem scales with ${\mathcal O}(\sum_{k=1}^{K}N_k^{3.5})$ if the interior point method is employed \cite{Luo2010}. Furthermore, the complexity of using the greedy search to update $\mathbf s$ scales with ${\mathcal{O}}(LKN^2)$. In summary, the overall complexity of Algorithm \ref{Algorithm3} under independent decoding scales with ${\mathcal O}(I_{\rm{AO}}(LKN^2+I_{\rm{MM}}\sum_{k=1}^{K}N_k^{3.5}))$, which is of a polynomial order.

\begin{algorithm}[!t]
  \caption{Proposed algorithm for solving problem \eqref{P_1}}
  \label{Algorithm3}
  \begin{algorithmic}[1]
    \STATE Initialize feasible $\left\{{\bm\Lambda}^{(0)},{\mathbf{s}}^{(0)}\right\}$ or $\left\{{\mathbf Q}^{(0)},{\mathbf{s}}^{(0)}\right\}$ and iteration index $t=0$
    \REPEAT
    \STATE Set $t=t+1$;
    \STATE Calculate the auxiliary parameters $\{{\bm\psi}^{(t)},{\bm\gamma}^{(t)}\}$ or $\{\hat{\bm\psi}^{(t)},\overline{\bm\psi}^{(t)},\hat{\bm\gamma}^{(t)},\overline{\bm\gamma}^{(t)}\}$ by Lemma \ref{Lemma_DE} or Lemma \ref{Lemma_DE2}
    \STATE Update ${\bm\Lambda}^{(t)}$ by \eqref{Power_Allocation_Design} or ${\mathbf Q}^{(t)}$ by Algorithm \ref{Algorithm2}
    \STATE Update ${\mathbf s}^{(t)}$ using the greedy search-based method
    \UNTIL{convergence}
  \end{algorithmic}
\end{algorithm}

\begin{figure*}[!t]
    \centering
    \subfigbottomskip=0pt
	\subfigcapskip=-5pt
\begin{tabular}{ccc}
        \begin{minipage}[t]{2.35in}
        \setlength{\abovecaptionskip}{0pt}
        \includegraphics[width=2.3in]{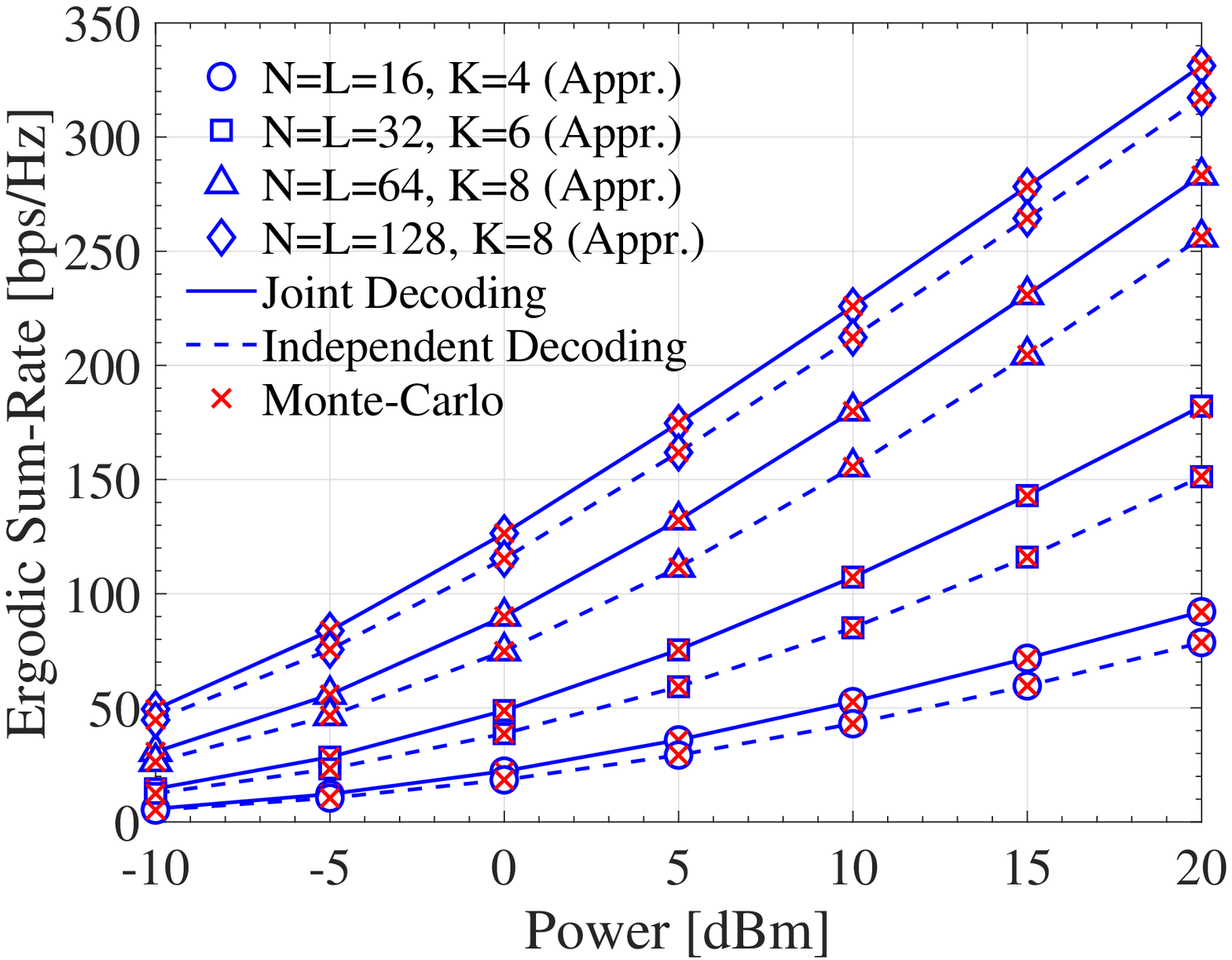}
        \caption{Asymptotic approximations.}
        \label{Figure1}
        \end{minipage}
        \hspace{-7pt}
        \begin{minipage}[t]{2.35in}
        \setlength{\abovecaptionskip}{0pt}
        \includegraphics[width=2.3in]{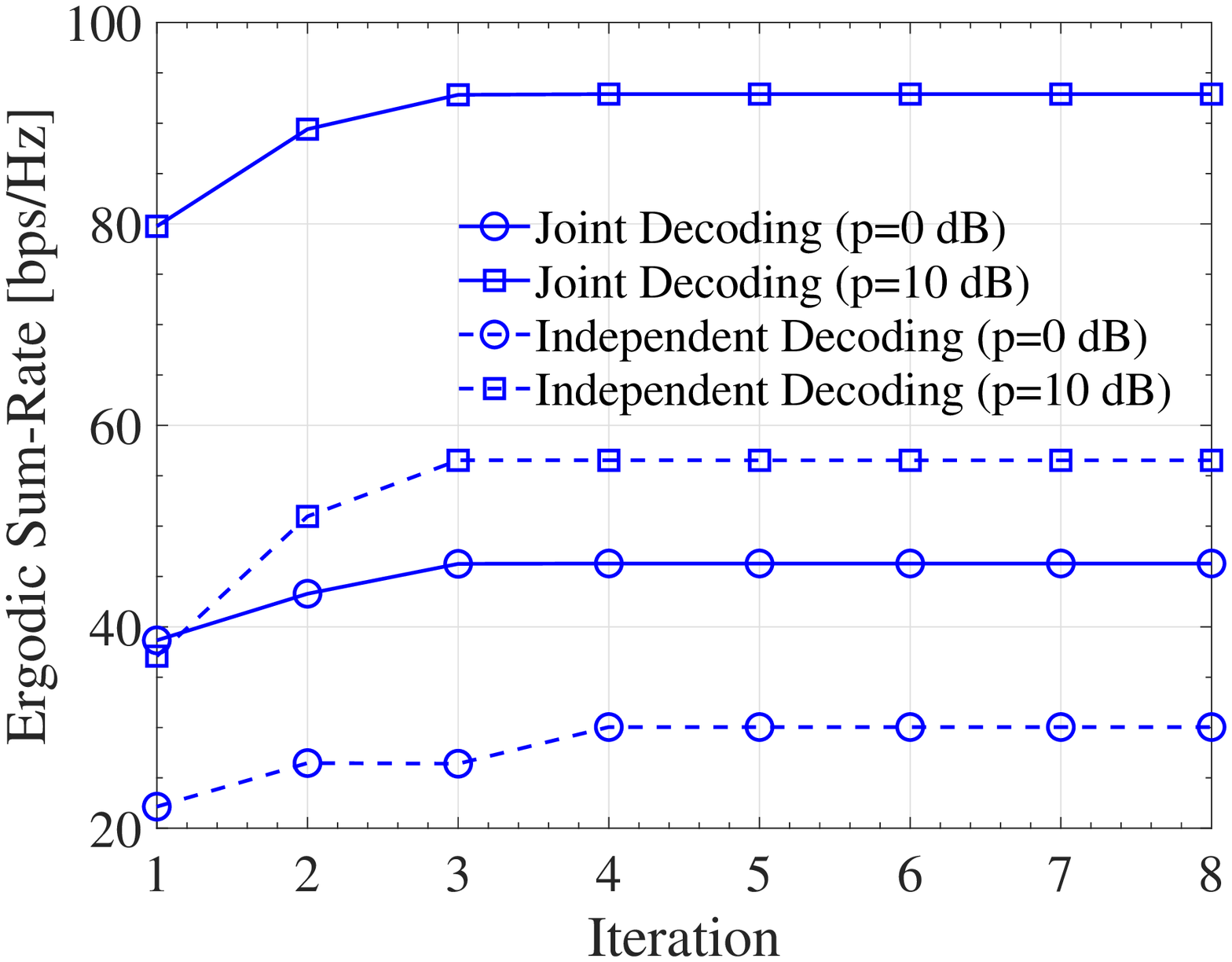}
        \caption{Convergence performance.}
        \label{Figure2}
        \end{minipage}
        \hspace{-7pt}
        \begin{minipage}[t]{2.35in}
        \setlength{\abovecaptionskip}{0pt}
        \includegraphics[width=2.3in]{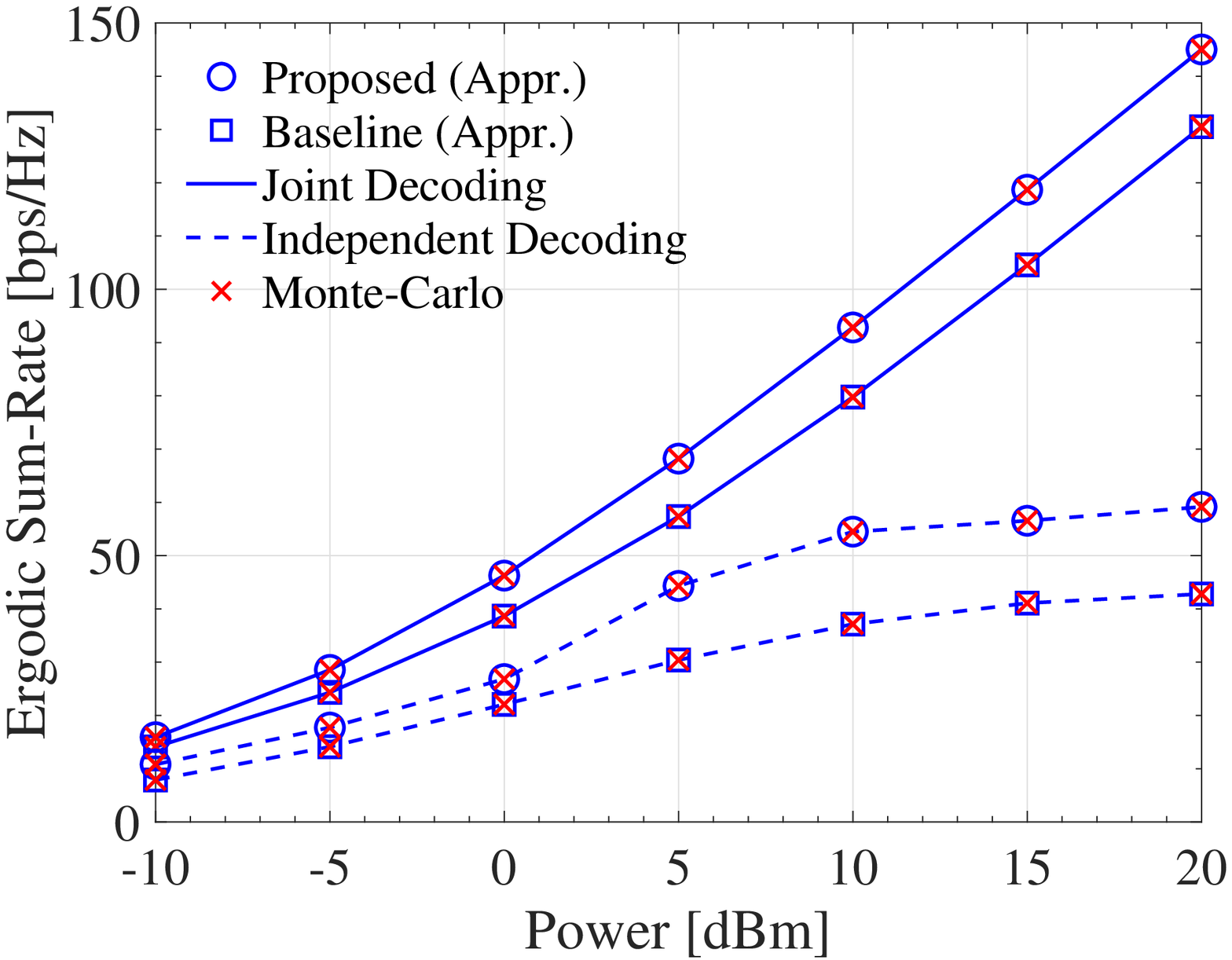}
        \caption{Sum-rate performance.}
        \label{Figure3}
        \end{minipage}
    \end{tabular}
    \vspace{-20pt}
\end{figure*}

\section{Numerical Results}
We now evaluate the performance of the proposed statistical-CSI-based scheme for uplink MU-MIMO communications through numerical simulations. We consider a scenario in which the UTs are distributed uniformly over a hexagonal cell. The path loss is set as $-120$ dB for all the UTs. The long-term statistics of the channels, i.e., , ${\mathcal{H}}_{\rm{s}}=\{{\mathbf U}_{{\rm R}_k},{\mathbf U}_{{\rm T}_k},{\bm\Omega}_k~{\text{for}}~k\in[K]\}$, are generated according to the methods in \cite{Lu2016}. Unless further specified, the simulation parameters are given as follows: $N=128$, $L=16$, $K=8$, $p_k=p$, $N_k=4$, for $k\in[K]$, and $\sigma^2=-120$ dBm. All the optimization variables are randomly initialized.

{\figurename} {\ref{Figure1}} verifies the accuracy of the asymptotic approximations derived in Lemma \ref{Lemma_DE} and Lemma \ref{Lemma_DE2}. As the figure shows, the asymptotic expressions closely track the computationally expensive Monte-Carlo simulations, even for rather small dimensions. The numerical results imply that the derived approximations accurately estimate the ergodic sum-rate in both cases. This confirms the validity of the proposed approach for joint precoding and antenna selection in the MU-MIMO uplink system with statistical CSI.

{\figurename} {\ref{Figure2}} plots the achievable ergodic sum-rate against the number of iterations in Algorithms \ref{Algorithm3}. The results imply that under both joint and independent decoding, the proposed algorithm quickly converges (usually converges after only three or four iterations). We can further observe that the joint decoding achieves a higher ergodic sum-rate than the independent decoding, which is consistent with the previous statements.

To further verify the throughput advantages brought by Algorithms \ref{Algorithm3}, we compare the sum-rate performance of our proposed joint design scheme and one baseline scheme in which $\mathbf s$ is randomly set and ${\mathbf Q}_k=\frac{p_k}{N_k}{\mathbf I}_{N_k}$ for $k\in[K]$. In {\figurename} {\ref{Figure3}}, the sum-rate is plotted against the transmit power budget $p$ for the above two schemes under both joint and independent decoding. One can observe that the proposed joint design significantly outperforms the baseline one in terms of the system throughput.

\section{Conclusion}
We proposed a joint antenna selection and precoding scheme for uplink transmission in MU-MIMO systems based on statistical CSI. Invoking tools from the random matrix theory, we derived closed-form expressions for the asymptotic ergodic sum-rate considering joint and independent decoding schemes at the receiver. The asymptotic terms were then used to develop an AO-based iterative algorithm for throughput maximization. Our numerical results imply that the proposed collaborative design considerably boosts the achievable throughput at a tractable computational complexity. As compared to the instantaneous-CSI-based scheme, it further requires a drastically lower update rate for the switching network and UT precoders. It is hence a good candidate for the use-cases in which wireless channels vary rather fast over time.


\begin{thebibliography}{00}
\bibitem{Molisch2004} A. F. Molisch and M. Z. Win, ``MIMO systems with antenna selection,'' \emph{IEEE Microw. Mag.}, vol. 5, no. 1, pp. 46--56, Mar. 2004.
\bibitem{Asaad2018} S. Asaad, A. M. Rabiei, and R. R. M\"{u}ller, ``Massive MIMO with antenna selection: Fundamental limits and applications,'' \emph{IEEE Trans. Wireless Commun.}, vol. 17, no. 12, pp. 8502--8516, Dec. 2018.
\bibitem{Gershman2004} M. Gharavi-Alkhansari and A. B. Gershman, ``Fast antenna subset selection in MIMO systems,'' \emph{IEEE Trans. Signal Process.}, vol. 52, no. 2, pp. 339--347, Feb. 2004.
\bibitem{Bereyhi2017} A. Bereyhi, M. A. Sedaghat, and R. R. M{\"u}ller, ``Asymptotics of nonlinear LSE precoders with applications to transmit antenna selection,'' in \emph{Proc. IEEE Int. Symp. Inf. Theory (ISIT)}, pp. 81--85, 2017.
\bibitem{Asaad2018_JSAC} S. Asaad \emph{et al.}, ``Optimal transmit antenna selection for massive MIMO wiretap channels,'' \emph{IEEE J. Sel. Areas Commun.}, vol. 36, no. 4, pp. 817--828, Apr. 2018.
\bibitem{Kuai2020} Z. Kuai and S. Wang, ``Thompson sampling-based antenna selection with partial CSI for TDD massive MIMO systems,'' \emph{IEEE Trans. Commun.}, vol. 68, no. 12, pp. 7533--7546, Dec. 2020.
\bibitem{Ouyang2020_TCOM} C. Ouyang \emph{et al.}, ``Receive antenna selection under discrete inputs: Approximation and applications,'' \emph{IEEE Trans. Commun.}, vol. 68, no. 4, pp. 2634--2647, Apr. 2020.
\bibitem{Gao2018} Y. Gao \emph{et al.}, ``Massive MIMO antenna selection: Switching architectures, capacity bounds, and optimal antenna selection algorithms,'' \emph{IEEE Trans. Signal Process.}, vol. 66, no. 5, pp. 1346--1360, Mar. 2018.
\bibitem{Ouyang2019} C. Ouyang \emph{et al.}, ``Optimal transmit antenna selection algorithm in massive MIMOME channels,'' in \emph{Proc. IEEE Wireless
Commun. Netw. Conf. (WCNC)}, pp. 1--6, 2019.
\bibitem{Ali2018} A. Bereyhi, S. Asaad, and R. R. M\"{u}ller, ``Stepwise transmit antenna selection in downlink massive multiuser {MIMO},'' \emph{Proc. 22nd Int. ITG Workshop Smart Antennas (WSA)}, pp. 1--8, 2018.
\bibitem{Dai2006} L. Dai, S. Sfar, and K. Letaief, ``Optimal antenna selection based on capacity maximization for MIMO systems in correlated channels,'' \emph{IEEE Trans. Commun.}, vol. 54, no. 3, pp. 563--573, Mar. 2006.
\bibitem{Mehta2021} R. Sarvendranath \emph{et al.}, ``Statistical CSI driven transmit antenna selection and power adaptation in underlay spectrum sharing systems,'' \emph{IEEE Trans. Commun.}, vol. 69, no. 5, pp. 2923--2934, May 2021.
\bibitem{Lu2022} J. Lu \emph{et al.}, ``Antenna selection for spatial correlated channel by exploiting statistical CSI,'' in \emph{Proc. IEEE/CIC Int. Conf. Commun. China (ICCC)}, pp. 1--5, 2022.
\bibitem{Lu2016} A.-A. Lu, X. Gao, and C. Xiao, ``Free deterministic equivalents for the analysis of MIMO multiple access channel,'' \emph{IEEE Trans. Inf. Theory}, vol. 62, no. 8, pp. 4604--4629, Aug. 2016.
\bibitem{Tulino2006} A. M. Tulino, A. Lozano, and S. Verdu, ``Capacity-achieving input covariance for single-user multi-antenna channels,'' \emph{IEEE Trans. Wireless Commun.}, vol. 5, no. 3, pp. 662--671, Mar. 2006.
\bibitem{Couillet2011} R. Couillet, M. Debbah, and J. W. Silverstein, ``A deterministic equivalent for the analysis of correlated MIMO multiple access channels,'' \emph{IEEE Trans. Inf. Theory}, vol. 57, pp. 3493--3514, June 2011.
\bibitem{Sun2017} Y. Sun, P. Babu, and D. P. Palomar, ``Majorization-minimization algorithms in signal processing, communications, and machine learning,'' \emph{IEEE Trans. Signal Process.}, vol. 65, no. 3, pp. 794--816, Feb. 2017.
\bibitem{Boyd2014} M. Grant and S. Boyd, ``CVX: Matlab software for disciplined convex programming, version 2.1,'' \url{http://cvxr.com/cvx}, Mar. 2014.
\bibitem{Luo2010} Z.-Q. Luo, W.-K. Ma, A. M.-C. So, Y. Ye, and S. Zhang, ``Semidefinite relaxation of quadratic optimization problems,'' \emph{IEEE Signal Process. Mag.}, vol. 27, no. 3, pp. 20--34, May 2010.
\end{thebibliography}
\end{document}